\documentclass[%
 reprint,
superscriptaddress,
 amsmath,amssymb,
 aps,
prl,
showkeys,
]{revtex4-2}

\usepackage{graphicx}
\usepackage{dcolumn}
\usepackage{bm}
\usepackage{array}
\usepackage{mathrsfs}
\usepackage{ulem}
\usepackage{hyperref}
\hypersetup{
  colorlinks = true,
  urlcolor = blue,
  linkcolor = blue,
  citecolor = green,
  filecolor = magenta,
}
\usepackage[sort&compress]{natbib}

\newcolumntype{M}{>{$}c<{$}}
\newcommand{\ud}{\mathrm{d}}
\newcommand{\pd}{\partial}

\begin{document}

\title{Space-borne Interferometers to Detect Thousands of Memory Signals Emitted by Stellar-mass Binary Black Holes}

\author{Shaoqi Hou}
\affiliation{School of Physics and Technology, Wuhan University, Wuhan, Hubei 430072, China}

\author{Zhi-Chao Zhao}
\email{zhaozc@cau.edu.cn}
\affiliation{Department of Applied Physics, College of Science, China Agricultural University, Qinghua East Road, Beijing 100083, People's Republic of China}

\author{Zhoujian Cao}
\affiliation{School of Physics and Astronomy, Beijing Normal University, Beijing 100875, China}

\author{Zong-Hong Zhu}
\email{zhuzh@whu.edu.cn}
\affiliation{School of Physics and Technology, Wuhan University, Wuhan, Hubei 430072, China}


\date{\today}

\begin{abstract}
    \textcolor{black}{The gravitational memory effect manifests gravitational nonlinearity, degenerate vacua, and asymptotic symmetries; its detection is considered challenging.
We propose using the space-borne interferometer to detect memory signals from stellar-mass binary black holes (BBHs), typically targeted by ground-based detectors. We use DECIGO detector as an example.
Over 5 years, DECIGO is estimated to detect $\sim$2,036 memory signals (SNRs $>$3) from stellar-mass BBHs.
Simulations used frequency-domain memory waveforms for direct SNR estimation.
Predictions utilized a GWTC-3 constrained BBH population model (Power Law + Peak mass, DEFAULT spin, Madau-Dickinson merger rate).
The analysis used conservative lower merger rate limits and considered orbital eccentricity.
The high detection rate stems from strong memory signals within DECIGO's bandwidth and the abundance of stellar-mass BBHs.
This substantial, conservative detection count enables statistical use of the memory effect for fundamental physics and astrophysics.
DECIGO exemplifies that space interferometers may better detect memory signals from smaller mass binaries than their typical targets.
Detectors in lower frequency bands are expected to find strong memory signals from $\sim 10^4 M_\odot$ binaries.}
\end{abstract}

\keywords{gravitational wave memory effects, space-borne interferometers, gravitational waves}

\maketitle

\section{Introduction}
The memory effect is a direct prediction of general relativity resulting from its nonlinearity \cite{Zeldovich:1974gvh,Braginsky:1986ia,Christodoulou1991,Thorne:1992sdb}.
It causes a permanent change in the spacetime metric, resulting in the arms of interferometers being stretched or contracted permanently.
It is closely connected to the Bondi-Metzner-Sachs (BMS) symmetries \cite{Sachs1962asgr} of isolated systems \cite{Flanagan:2015pxa,Strominger:2018inf}.
As a form of BMS symmetry, supertranslations transform degenerate gravitational vacuum states into one another, manifesting as the memory effect \cite{Ashtekar:1981hw,Strominger2014bms,Strominger:2014pwa}. The null energy flux, which is the Noether charge conjugate to supertranslation, serves as the source of the (nonlinear) memory effect \cite{Strominger:2014pwa,Flanagan:2015pxa}.
Recently discovered spin and center-of-mass (CM) memories  \cite{Pasterski:2015tva,Nichols:2017rqr,Nichols:2018qac} demand the enlargement of the BMS group \cite{Barnich:2009se,Barnich:2010eb,Campiglia:2014yka,Campiglia:2015yka,Campiglia:2020qvc}.
To differentiate them, the earlier memory effect is specifically referred to as displacement memory. Generally, displacement memory is the strongest, while CM memory is the weakest \cite{Nichols:2017rqr,Nichols:2018qac}.
In this work, we will focus on the displacement memory, referred to simply as \textit{the} memory effect unless otherwise specified.

There are several intriguing applications of memory effects. For instance, the memory effect can be used to test the consistency between different waveform models \cite{Khera:2020mcz}. It may also help distinguish neutron star-black hole (NS-BH) mergers from binary black hole (BBH) mergers \cite{Tiwari:2021gfl}, and by considering matter effects, it could differentiate binary neutron star (BNS) mergers from NS-BH mergers as well \cite{Lopez:2023aja}.
Incorporating nonlinear memory waveforms in parameter estimation might break the degeneracy between the inclination angle $\iota$ and the luminosity distance $D_L$, especially for equal-mass, non-precessing binaries \cite{Xu:2024ybt}.
Additionally, it has been suggested that observing specific memory effects could reveal the corresponding asymptotic symmetries \cite{Goncharov:2023woe}.
Studies on modified gravity theories proposed using the memory effect to probe the nature of gravity \cite{Hou:2020tnd,Tahura:2020vsa,Seraj:2021qja,Hou:2020wbo,Tahura:2021hbk,Hou:2021bxz,Heisenberg:2023prj,Hou:2023pfz}. Furthermore, since the memory effect does not manifest in odd spacetime dimensions, detecting it would rule out such higher-dimensional models \cite{Hollands:2004ac,Hollands:2016oma}.

However, due to the signal's weakness, no memory effect has been detected by interferometers \cite{Lasky:2016knh,Yang:2018ceq,Hubner:2019sly,Zhao:2021hmx,Islam:2021old,Hubner:2021amk,Cheung:2024zow} or pulsar timing arrays \cite{NANOGrav:2023vfo}.
Recently, the North American Nanohertz Observatory for Gravitational Waves (NANOGrav) set a median upper limit of approximately $3.3\times10^{-14}$ on the strain of the memory effect \cite{NANOGrav:2023vfo}.
To date, studies have predicted that detecting the memory effect is challenging.
For example, networks of the LVK detectors and LIGO India may accumulate sufficient {signal-to-noise ratios (SNRs), at least 3}, for the memory effect after 3 to 5 years of observing hundreds or thousands of stellar-mass BBH merger events \cite{Hubner:2019sly,Grant:2022bla}.
A network of 2 or 3 Cosmic Explorers (CEs) could detect memory effects from individual loud events, with a yearly detection rate of 4 to 7 \cite{Grant:2022bla}.
TianQin is expected to detect around 0.5 to 2 memory signals {(SNRs $\geq$ 3)} during its 5-year operation \cite{Sun:2022pvh}.
LISA, on the other hand, is predicted to observe tens to hundreds of memory signals {(SNRs $\geq$ 1)} from massive BBHs over its 4-year operation. If {the SNR threshold is set to 5}, a few to tens of signals may be observable \cite{Gasparotto:2023fcg,Inchauspe:2024ibs}.

Although these predictions hold promise for detecting this interesting phenomenon, the relatively low rates may limit the practical applications of memory effects.
One shall realize that in the previous works \cite{Hubner:2019sly,Grant:2022bla,Sun:2022pvh,Gasparotto:2023fcg,Inchauspe:2024ibs}, people conventionally considered the memory effect produced by the very binaries chirping at the bandwidths of the respective interferometers.
However, the memory signal may actually be strong at relatively lower frequency ranges as suggested by Ref.~\cite{Ghosh:2023rbe}.
So in this work, we proposed to use the space-borne interferometer to detect the memory effect generated by stellar-mass BBHs, which are abundant and usually thought to be the targets of the ground-based interferometers.
We will consider the DECihertz laser Interferometer Gravitational wave Observatory (DECIGO) \cite{Seto:2001qf,decigo2019,Kawamura:2020pcg} as one example \footnote{Indeed, the downscaled version, B-DECIGO, was also expected to observe a lot of memory signals, qualitatively \cite{Ando:report}.}.
The similar idea can be applied to LISA \cite{Audley:2017drz}, Taiji \cite{Taiji2017} or TianQin \cite{Luo:2015ght} with the caution that a little bit heavier binaries (on the order of $10^2M_\odot-10^3M_\odot$) may produce loud enough memory signals for them.
We will show that the memory signals produced by these sources are sufficiently loud for DECIGO, due to its low frequency nature, and so the detect rates are substantially high.
Moreover, the stellar-mass BBHs have been confirmed to exist by the observations of the LVK collaboration, allowing for more reliable estimates of their coalescence rates.

\section{Methods}
\subsection{Memory waveform}
\label{sec-mem}

The memory effect $h_\text{D}$ can be effectively calculated using the flux-balance laws \cite{Flanagan:2015pxa,Mitman:2020pbt} in the Bondi-Sachs formalism with the coordinates $(u,r,\theta,\phi)$ \cite{Bondi:1962px,Sachs:1962wk},
\begin{equation}
	\label{eq-hd}
    \begin{aligned}
	h_\text{D}=&r\widetilde\sum(-1)^{\tilde m }{}_{-2}Y_{\ell m}\\&\times\sqrt{\frac{(\ell-2)!}{(\ell+2)!}}\mathcal C_\ell(-2,\hat \ell ,\hat m ;2,\tilde \ell ,-\tilde m )
	\int_{u_0}^u\dot h_{\hat \ell \hat m }\dot{ h}^*_{\tilde \ell \tilde m }\ud u'.
    \end{aligned}
\end{equation}
Here, in this formula, ${}_{-2}Y_{\ell m}(\theta,\phi)$ is the spin-weighted spherical harmonics, $h_{\ell m}$ is the spherical mode of the complex strain $h=h_+-ih_\times=\sum_{\ell m}{}_{-2}Y_{\ell m}h_{\ell m}$, dot means $\pd_u$, star means to take the complex conjugate, and the sum $\widetilde\sum$ runs over $(\ell,m)$, $(\hat \ell , \hat m )$ and $(\tilde \ell , \tilde m )$.
One also defines \cite{Nichols:2017rqr}
\begin{equation*}
	\label{eq-def-cl}
	\mathcal C_\ell(\hat s,\hat \ell ,\hat m ;\tilde s,\tilde \ell ,\tilde m )
	\equiv\int\ud^2\theta\sin\theta({}_{\hat s}Y_{\hat \ell \hat m })({}_{\tilde s}Y_{\tilde \ell \tilde m })({}_{s}\bar Y_{\ell m}),
\end{equation*}
which vanishes unless $s=\hat s+\tilde s$, $m=\hat m +\tilde m $, and $\text{max}\{|\hat \ell -\tilde \ell|,|\hat m+\tilde m |,|\hat s+\tilde s|\}\le\ell\le \hat\ell+\tilde \ell $.
In Eq.~\eqref{eq-hd}, one ignores the so-called linear memory effect, which is smaller than the nonlinear one given by this expression \cite{Mitman:2020pbt}.
The $+$- and $\times$-components are $\Re h_\text{D}$ and $-\Im h_\text{D}$, respectively.
As is well-known, for non-precessing binary systems, the memory waveform is mainly given by the terms with $\ell=2,4$ and $m=0$ in Eq.~\eqref{eq-hd} \cite{Favata:2008yd,Favata:2009ii,Favata:2010zu,Favata:2011qi}, i.e.,
\begin{equation}
	\label{eq-hmin}
	h_\text{D}
	\approx\frac{(17+\cos^2\iota)\sin^2\iota}{384\pi}r
	\int_{u_0}^{u}\ud u'\int\ud^2\theta\sin\theta(\dot h_+^2+\dot h_\times^2),
\end{equation}
where $\iota$ is the inclination angle.
This results in the \textit{minimal-waveform} model \cite{Favata:2010zu}.
In our simulation, we did not use this model, but instead used the Fourier transform of the full expression \eqref{eq-hd} to calculate the memory waveform.
Despite this, it would be still interesting to know the memory waveform generated by the oscillatory mode $(2,\pm2)$ in the early inspiral phase,
\begin{equation}
	\label{eq-hd-in22}
	h_\text{D}=\frac{\mathcal M(\pi\mathcal Mf)^{2/3}}{48r}\sin^2\iota(17+\cos^2\iota),
\end{equation}
where $\mathcal M$ is the chirp mass and $f$ is the GW frequency of the corresponding oscillatory $(2,\pm2)$ modes.

The frequency-domain memory waveform is given by,
\begin{equation}
	\label{eq-fd-wf}
    \begin{aligned}
	\tilde h_\text{D}=&\frac{2\pi r}{if}\widetilde\sum(-1)^{\tilde m }{}_{-2}Y_{\ell m}\sqrt{\frac{(\ell-2)!}{(\ell+2)!}}\mathcal C_\ell(-2,\hat \ell ,\hat m ;2,\tilde \ell ,-\tilde m ) \\
	&\times \int f'(f'-f)\tilde h_{\hat \ell \hat m }(f')\tilde h^*_{\tilde \ell \tilde m }(f'-f)\ud f',
    \end{aligned}
\end{equation}
where $\tilde h_{\ell m}(f)$ is the Fourier transform
\begin{equation}
	\tilde h_{\ell m}(f)=\int_{-\infty}^{+\infty}h_{\ell m}(u)e^{-i2\pi fu}\ud u.
\end{equation}
Note the existence of $f$ in the denominator of Eq.~\eqref{eq-fd-wf}.
This is one of the key characteristics of the frequency-domain memory waveform.
In order to get this result, one makes use of the following trick \cite{Schwartz:2013pla},
\begin{equation}
    \begin{aligned}
	\int_{-\infty}^{+\infty}&\ud ue^{-i\omega u}\int_{-\infty}^u\ud u'e^{i(\omega_1-\omega_2)u'} \\
	&=  \int_{-\infty}^{+\infty}\ud ue^{-i\omega u}\int_{-\infty}^{+\infty}\ud u'\Theta(u-u')e^{i(\omega_1-\omega_2)u'}\\
	&=  \frac{2\pi}{i\omega}\delta(\omega_1-\omega_2-\omega),
    \end{aligned}
\end{equation}
with $\Theta(x)$ the Heaviside function.
Equation~\eqref{eq-fd-wf} allows one to compute the frequency-domain memory waveform by substituting the corresponding oscillatory waveform in the frequency domain.
Using Eq.~\eqref{eq-fd-wf}, one can obtain the memory waveform sourced by the oscillatory modes in the inspiral phase.
Since we used IMRPhenomXPHM \cite{Pratten:2020ceb} to generate the oscillatory modes, we need determine the upper frequency bound $f_\text{max}$ at which the inspiral phase ends.
According to Refs.~\cite{Garcia-Quiros:2020qpx,Pratten:2020fqn,Pratten:2020ceb}, the inspiral phase ends at different frequencies for the phase and the amplitude.
These frequencies also depend on the magnetic quantum number $m$ [See, e.g., Eq.~(4.4) in Ref.~\cite{Garcia-Quiros:2020qpx}].
But it can be checked that for a fixed $m$, the frequency bounds for the phase and the amplitude are very close to each other.
And if one multiplies these frequency bounds by a factor of $2/m$, they are also very similar.
Therefore, in our simulation, we computed all the frequency bounds, multiplied them by $2/m$, and then took the minimal one as $f_\text{max}$.
{\color{black} See Fig.~\ref{fig-wftd} for a representative time-domain example (oscillatory plus memory), and Fig.~\ref{fig-wftdiotas} for complementary illustrations discussed in this subsection.
}

To estimate the memory SNR, we actually used the frequency-domain waveform $\tilde h_\text{D}$ given by Eq.~\eqref{eq-fd-wf}.
We did not calculate the time-domain waveform $h_\text{D}$, and then Fourier transform it to get $\tilde h_\text{D}$.
In order to perform the Fourier transformation, one has to place a clever window function to minimize the Gibbs effect.
The window function shall be chosen carefully on an event basis.
The direct application of Eq.~\eqref{eq-fd-wf} avoids this issue.
Of course, any detector operates for a finite time, but we considered that DECIGO works for $T=5$ yrs.
This time period is much larger than the time scale of the memory effect produced by stellar-mass BBHs.
So it is appropriate to directly use the Fourier transform $\tilde h_\text{D}$ of the memory waveform.
In order to get the memory SNR as accurate as possible, one shall use \verb+PyCBC+ to generate the oscillatory modes starting at a low enough frequency bound \verb+f_lower+.
It turned out that \verb+f_lower = 0.1+ Hz is a good choice.
Generally speaking, integrating Eq.~\eqref{eq-fd-wf} is very time-consuming.
Fortunately, the frequency dependence of $\tilde h_\text{D}$ is very simple, especially at low frequencies.
Even if at higher frequencies for large inclination angles, $\tilde h_\text{D}$ is more complicated \footnote{Refer to Fig.~\ref{fig-wftdiotas}.}, DECIGO is not sensitive at these frequencies.
So in the simulation, one can compute the memory waveform $\tilde h_\text{D}$ for a (relatively) small number of frequencies, and then use the interpolation method to get the values at other frequencies.
We tested that it is sufficient to sample $\tilde h_\text{D}$ at 100 frequencies, which are log-uniformly distributed in the range $[0.01, 100]$ Hz.
Increasing the number of frequencies up to 1000 changes the SNR by around 0.01\%.
However, this does not imply that the step size \verb+delta_f+ of the frequency for the oscillatory modes can be large, too.
It shall be small, which was set to be 0.001 Hz in our simulation.

\subsection{The noise power spectrum density}
\label{sec-psd}

DECIGO is a gravitational wave detector composed of four copies of constellations of three drag-free spacecraft \cite{Seto:2001qf,decigo2019,Kawamura:2020pcg}.
In each constellation, the spacecraft are separated from each other by about 1, 000 km, and they orbit around the Sun with a period of 1 yr.
Each constellation can be equivalently viewed as two uncorrelated, L-shaped detectors rotated by $45^\circ$.
The one-sided noise power spectrum of an L-shaped interferometer can be written as \cite{Kawamura:2011zz,Yagi:2011wg}:
\begin{equation}
\begin{aligned}
	\bar S_n(f)  =&  7.05 \times 10^{-48}\left[1+\left(\frac{f}{f_p}\right)^2\right]
	+4.8\times 10^{-51}\\&\times\frac{(1 \text{ Hz}/f)^4}{1+(f/f_p)^2}
	+5.53 \times 10^{-52}\left(\frac{1 \text{ Hz}}{f}\right)^{4} \mathrm{~Hz}^{-1},
\end{aligned}
\end{equation}
with $f_p=7.36$ Hz.
The galactic and the extragalactic compact binaries in the foreground would produce GWs that contaminate the sensitivity.
The effective noise power spectra due to these GWs are given by \cite{Nelemans:2001hp,Farmer:2003pa},
\begin{equation}
	\begin{aligned}
		S_n^{\text {g }}(f)  & =2.1 \times 10^{-45}\left(\frac{1 \rm ~H z}{f}\right)^{7 / 3} \rm Hz^{-1}, \\
		S_n^{\text {eg }}(f) & =4.2 \times 10^{-47}\left(\frac{1 \rm ~H z}{f}\right)^{7 / 3} \rm Hz^{-1},
	\end{aligned}
\end{equation}
respectively.
These spectra  should be further multiplied by a factor corresponding to the high-frequency cutoff{\color{black}, which models the roll-off of unresolved WD binaries}, $$F(f)=\exp\left[-2\left(\frac{f}{0.05{\rm ~Hz}}\right)^2\right].$$
The noise from neutron stars (NS) should also be considered.
However, following \cite{Yagi:2009zz}, we assume most (99\%) of them can be removed by foreground subtraction. Then, the noise is
\begin{equation}
	S'^{\rm NS}_n(f)  =0.01\times S_n^{\rm N S}(f)
	=1.3 \times 10^{-48}\left(\frac{1\rm ~H z}{f}\right)^{7 / 3}\rm H z^{-1}.
\end{equation}
The overall noise power is
\begin{equation}\label{eq:psd}
\begin{aligned}
	S_n(f) =
	&S_n^{\text {eg}}(f) F(f) + S'^{\rm NS}_n(f)\\&+ {\rm min}\bigg\{ \bar S_n(f)
	+S_n^{\rm g}(f)F(f), \frac{\bar S_n(f)}{\exp(-\kappa T^{-1}_{\rm obs}\frac{d N}{d f})}
	\bigg\}.
\end{aligned}
\end{equation}
In this expression, the symbol $\rm min$ implies to take the smaller argument.
$T_{\rm obs}$ is the observation time, which is set to 5 yrs.
$\ud N/\ud f$ is the number density of white dwarfs in the galaxy per unit frequency,
$$\frac{d N}{d f}=2\times10^{-3}\left(\frac{1\rm ~ Hz}{f}\right)^{11/3}{\rm ~Hz}^{-1}.$$
And finally, the factor appearing in the exponential function is $\kappa=4.5$ \cite{Sun:2023eic}.
The noise spectrum Eq.~\eqref{eq:psd} has been plotted in Fig.~\ref{fig-wftdiotas}. 

With Eq.~\eqref{eq:psd}, one can calculate the signal-to-noise ratio (SNR) for the memory waveform with
\begin{equation}
	\label{eq-snr2}
	\rho^2=4 \times{8} \int_0^{\infty} \frac{|\tilde{h}_\text{D}(f)|^2K(f)}{S_n(f)} \ud f.
\end{equation}
Here, the factor $K(f)$ is given by \cite{Zhang:2020khm},
\begin{equation}
	K(f) = \frac{3}{10} \frac{1}{1+\left(f / f_*\right)^2 /\left[1.85-0.58 \cos \left(2 f / f_*\right)\right]},
\end{equation}
where $f_*=1/2\pi L$ with $L=1000 \mathrm{~km}$ the arm length of the DECIGO detector \cite{Kawamura:2006up}.
This factor  is actually the averaged detector response function, and it comes from the fact that the four constellations of three satellites of DECIGO are constantly orbiting around the Sun, so the source orientation relative to each constellation plane varies over time.
For our purpose of estimating the detection rate of memory signals, one can average it over the source orientation and the  GW polarization angle.
In the end, the factor ${8}$ of Eq.~\eqref{eq-snr2} arises from 8 uncorrelated interferometric signals, which are collected by the four independent constellations, having similar orbital configurations and sharing the same instrumental noise PSD \cite{Kawamura:2011zz,Balcerzak:2012bv}

\subsection{Sampling binary black hole systems}
\label{sec-sbbh}

In our simulation, we generated BBHs with randomly assigned right ascension (RA), declination (DEC), inclination angle $\iota$, and coalescence phase $\phi$.
The binary components follow quasi-circular precessing orbits.
Their masses, spins, and merger rates were sampled from some relevant population synthesis models.
Specifically, we used the Power Law + Peak mass model {\color{black} to sample $(m_1,m_2)$ jointly} and the DEFAULT spin model to characterize the {\color{black} spin} distributions, with the results of Ref.~\cite{KAGRA:2021duu} serving as fiducial parameters.
For the evolution of the merger rate with redshift, given that the BBH detection horizon is smaller than the expected peak redshift $z_{\rm p}$, where the merger rate is maximized, the LVK detectors were unable to effectively model the  merger rate at high redshifts. {\color{black} We did not impose any intrinsic mass--redshift correlation, as the apparent trend of heavier systems at larger distances reflects selection rather than population evolution \cite{KAGRA:2021duu}.}
Therefore, we adopted the LVK-constrained local merger rate, which falls within the range of $17.9\operatorname{Gpc}^{-3}\operatorname{yr}^{-1}$ to $44\operatorname{Gpc}^{-3}\operatorname{yr}^{-1}$ at $z=0.2$ at the $90\%$ credible interval (C.L.).
We further assumed that the merger rate is proportional to the Madau–Dickinson star formation rate model \cite{Madau:2014bja}, which is consistent with the low-redshift merger rate inferred by LVK \cite{KAGRA:2021duu}.
With $R_0=17.9\operatorname{Gpc}^{-3}\operatorname{yr}^{-1}$, we estimated that there can be around $5\times10^4$ merger events per year.
The distributions of several binary parameters are shown in Fig.~\ref{fig:EventRate}.

\section{Results}

\subsection{Memory waveform characteristics}
\label{sec-mwf-c}

\begin{figure}[h]
	\centering
	\includegraphics[width=0.49\textwidth]{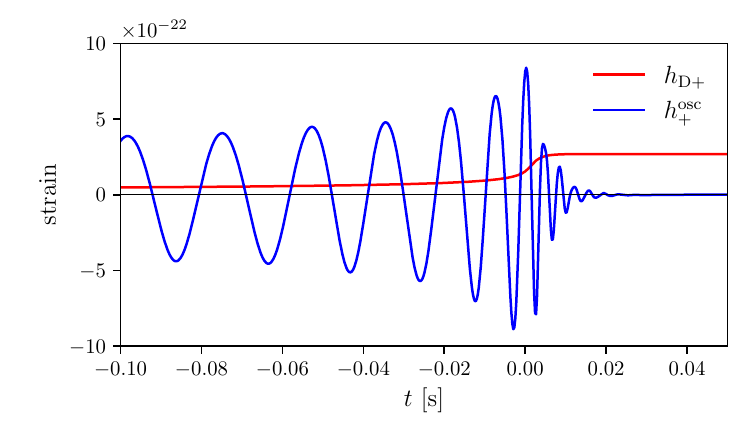}
	\caption{Time-domain waveforms produced by PyCBC from a GW150914-like binary system.
		The blue curve represents the $+$-polarization of the oscillatory mode $h_+^\text{osc}$ generated with IMRPhenomXPHM \cite{Pratten:2020ceb}, while the red curve illustrates the memory waveform $h_\text{D}$ calculated with Eq.~\eqref{eq-hd}. The inclination angle is set to $\iota = \pi/2$.}
	\label{fig-wftd}
\end{figure}

The memory waveform can be computed using the flux-balance laws based on the Bondi-Sachs formalism \cite{Flanagan:2015pxa,Mitman:2020bjf}, as briefly described in Section~\ref{sec-mem}.
With \verb+PyCBC+ \cite{alex_nitz_2024_10473621}, one can obtain the time-domain waveform according to Eq.~\eqref{eq-hd}.
Figure~\ref{fig-wftd} shows the plus polarization $h_\text{D+}$ of the memory waveform  generated by a  spinless binary system like GW150914 with masses $m_1=36M_\odot,m_2=30M_\odot$ at the distance $D_L=410$ Mpc, alongside the corresponding oscillatory mode $h_+^\text{osc}$.
The inclination angle is set to $\iota = \pi/2$, at which the memory amplitude determined by Eq.~\eqref{eq-hmin} is maximized.
To plot this figure, we used the waveform model IMRPhenomXPHM \cite{Pratten:2020ceb} to generate all of the available oscillatory modes, $(2,\pm1),(2,\pm2),(3,\pm2),(3,\pm3)$, and $(4,\pm4)$, and then calculated the memory waveform with Eq.~\eqref{eq-hd}.
For the chosen oscillatory modes, the spherical modes for the memory waveform, labeled by $\ell m$ in Eq.~\eqref{eq-hd}, are limited, so we considered all of them.
It can be shown that the plus($+$)-polarization $h_\text{D+}$ of the memory waveform is still much larger than the cross($\times$)-polarization $h_{\text{D}\times}$, which is not shown in Fig.~\ref{fig-wftd}.

\begin{figure*}
	\centering
	\includegraphics[width=0.9\textwidth]{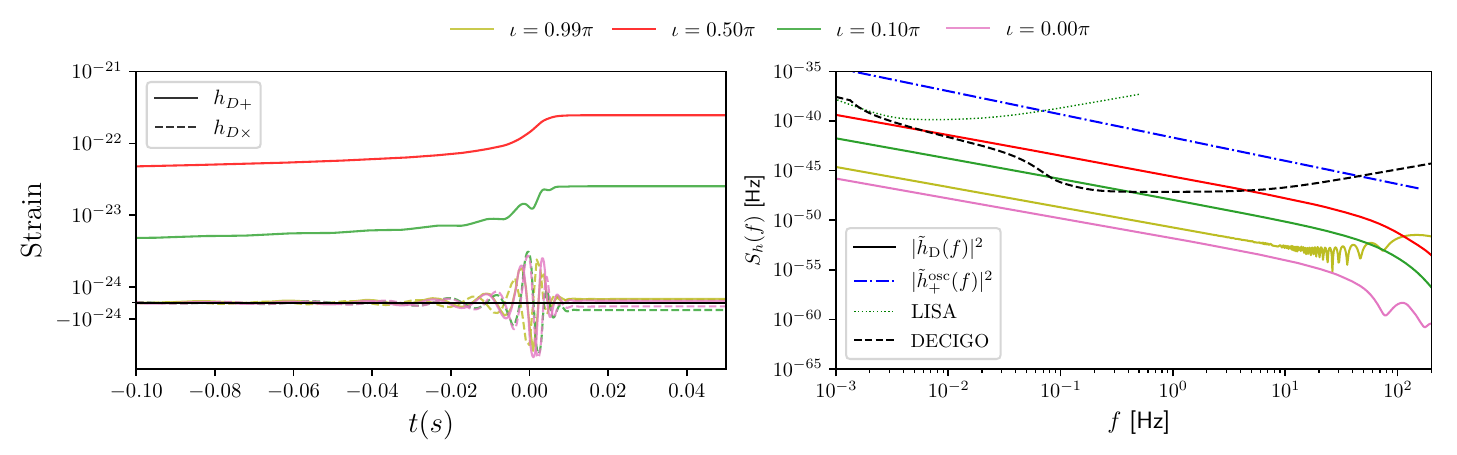}
	\caption{Left: $h_\text{D+}$ and $h_{\text{D}\times}$ as functions of the inclination angle $\iota$ for a GW150914-like binary system.
		Right: the Fourier transformed waveforms $|\tilde h_+^\text{osc}|^2$ and $|\tilde h_\text{D}|^2$ for different values of $\iota$, together with the sensitivity curves of DECIGO and LISA. 
        It should be noted that in the left panel, the solid lines corresponding to $\iota = 0.99\pi$ and $\iota = 0.00\pi$ almost coincide, so they appear as a single curve. 
	}
	\label{fig-wftdiotas}
\end{figure*}

At different inclination angles ($\iota$), $|h_\text{D+}|$ becomes smaller, while $|h_{\text{D}\times}|$ increases as shown in the left panel of Fig.~\ref{fig-wftdiotas} \footnote{Note that $|h_{\text{D}\times}$ at $\iota=\pi/2$ is so small that it is basically invisible.}.
$h_\text{D+}$ seems to be an even function of $\iota$, while $h_{\text{D}\times}$ is an odd function.
One may check this phenomenon by plotting more memory waveforms at different $\iota$'s, but this would make the figure too cluttered.
In particular, near $\iota=0$ and $\pi$, $h_\text{D+}$ oscillates while increasing.
On the contrary, $h_{\text{D}\times}$ always oscillates, and it may become positive or negative eventually.
Except at the $\iota$'s very close to $0$ or $\pi$, $|h_\text{D+}|$ is always larger than $|h_{\text{D}\times}|$.
These general features remain unchanged for spinning systems.
It shall be emphasized that when higher oscillatory modes are included, and $\iota\ne\pi/2$, the cross polarization $h_{\text{D}\times}$ exists \cite{Talbot:2018sgr}.

The corresponding frequency-domain memory waveforms could be obtained with Eq.~\eqref{eq-fd-wf}.
In Fig.~\ref{fig-wftdiotas}, the right panel displays the Fourier-transformed waveforms.
The blue dot-dashed curve is for $|\tilde h_+^\text{osc}|^2$, corresponding to the blue solid-curve in Fig.~\ref{fig-wftd}.
The remaining solid curves are $|\tilde h_\text{D}|^2$ at different $\iota$'s.
The noise power spectral densities (PSD's) of DECIGO (dashed) and the analytical sensitivity curve for LISA (dotted), as described in the Science Requirement Document \cite{lisa2018lisa,Babak:2021mhe}, are also plotted.
In calculating the PSD of DECIGO, we have accounted for the contamination from foreground galactic and extra-galactic compact binaries, as well as neutron stars \cite{Nelemans:2001hp,Farmer:2003pa,Yagi:2009zz,Kawamura:2011zz,Yagi:2011wg,Sun:2023eic}.
For a detailed construction of the PSD and the computation of the SNR, please refer to Section~\ref{sec-psd}.
Clearly,  the red curve for the memory effect is well above the PSD of DECIGO at the certain frequency range.
It turned out that the SNR of $\tilde h_\text{D}$ is $\sim 47.2$.
Therefore, even though a stellar-mass BBH chirps around $O(100)$ Hz, its memory signal could still be detected by DECIGO.
However, LISA's sensitivity curve (dotted) is above the memory signals produced by these GW150914-like events, so these kind of signals are unlikely to be observed by LISA.
On the right panel, $|\tilde h_\text{D}|^2$ is the largest at $\iota=\pi/2$ for the most frequencies.
It develops certain oscillating features at small $\iota$'s at high frequencies.
Note that in Fig.~\ref{fig-wftdiotas}, $|\tilde h_\text{D}|^2$ is plotted, instead of $|\tilde h_\text{D+}|^2$ or $|\tilde h_{\text{D}\times}|^2$, because $|\tilde h_\text{D}|^2$ is used to estimate the SNR.

\subsection{Sampled binary black hole systems}
\label{sec-sd-bbh}

\begin{figure}[h]
	\centering
	\includegraphics[width=0.95\linewidth]{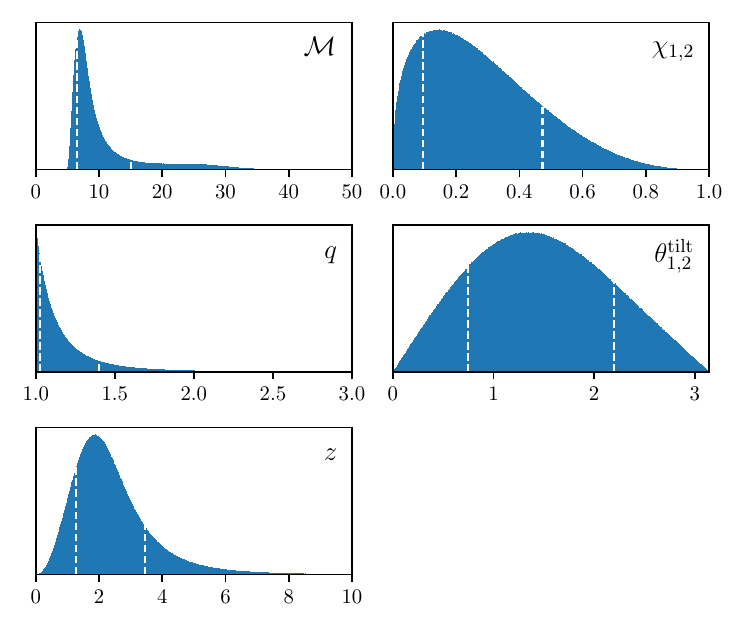}
	\caption{The distributions of the source-frame chirp mass $\mathcal{M}$, the dimensionless spins $\chi_{1,2}$, the mass ratio $q$, the tilt angles $\theta_{1,2}^\text{tilt}$, and the redshift $z$, sampled from binary black hole population model obtained by LVK using GWTC-3. The vertical dashed lines indicate the $1\sigma$ confidence level. 
	}
	\label{fig:EventRate}
\end{figure}

We sampled the BBH systems with the population model determined according to the most recent GW observations in GWTC-3 \cite{KAGRA:2021duu}.
The distributions of the chirp mass $\mathcal{M}$, the dimensionless spins $\chi_{1,2}$, the mass ratio $q=m_1 / m_2 $, the tilt angles $\theta_{1,2}^\text{tilt}$, and the redshift $z$ are displayed in Fig.~\ref{fig:EventRate}.
As shown, most of the sampled BBH systems have small chirp masses, with $\mathcal{M} \sim 8 M_\odot$, $q \sim 1$, and $z\sim 2$.
Lower redshifts imply louder signals, enabling the easier detection.
In addition, more symmetrical binaries generate stronger memory signals, at least in the inspiral stage, refering to Eq.~\eqref{eq-hd-in22} and Refs.~\cite{Favata:2008ti,Favata:2008yd,Favata:2009ii,Favata:2011qi}.
A lot of stars have the dimensionless spins $\lesssim0.4$, and the spin's tilt angles distribute quite symmetrically about $\pi/2$.

\subsection{Event rates}
\label{sec-rate}

Using the sampled BBHs, we generated their frequency-domain memory waveforms based on Eq.~\eqref{eq-fd-wf}.
The oscillatory components were obtained using \verb+PyCBC+ \cite{alex_nitz_2024_10473621} with the IMRPhenomXPHM model \cite{Pratten:2020ceb}.
All available spherical modes of the oscillatory components were taken into account, and we also computed all possible spherical modes of the memory signal.

To claim the detection of a specific GW signal, it is generally required that $\rho \ge 8$ \cite{Isoyama:2018rjb,Gerosa:2019dbe}.
However, different thresholds, such as $\rho \ge 10$ and $\rho \ge 5$, have also been employed in some cases \cite{Timpano:2005gm,Katz:2018dgn,Gourgoulhon:2019iyu,Babak:2023lro}.
For memory signals, a fourth threshold of $\rho \ge 3$ was advocated in Refs.~\cite{Lasky:2016knh,Johnson:2018xly,Grant:2022bla}.
If one views the memory waveform as a non-oscillatory residual on top of the oscillatory component, one may require $\rho \ge 1$ in order to claim the detection of the memory signal \cite{Lindblom:2008cm,Hu:2022rjq}.
So Table~\ref{tab:Results} lists the numbers of detectable memory signals during DECIGO's 5-year observation period, meeting the different SNR thresholds.
\begin{table}[h]
	\centering
	\setlength{\tabcolsep}{.5em}
	\begin{tabular}{ c c c c c c}
		\hline
		\hline
		SNR Thresholds & 10       & 8         & 5         & 3          & 1            \\
		\hline
		Event Numbers  & $\sim53$ & $\sim120$ & $\sim533$ & $\sim2015$ & $\sim 15364$ \\
		\hline
	\end{tabular}
	\caption{The numbers of memory signals with sufficiently high SNRs that might be detected by DECIGO during its 5-year operation, assuming $R_0=17.9\text{ Gpc}^{-3}\text{yr}^{-1}$.
		Note that if the upper limit were used, the event counts would be increased by a factor of 1.5.
	}
	\label{tab:Results}
\end{table}
It shows that up to 2015  memory signals with SNRs $> 3$ could be observed by DECIGO.
Or, if one raises the SNR threshold to 8, the number of detectable memory signals would be 120.
We also found that among the 120 loudest events (the second column), 5 of them have SNRs, accumulated \textit{solely} in the inspiral stage, greater than 3.
Note that $R_0=17.9\text{ Gpc}^{-3}\text{yr}^{-1}$(90\% C.L. lower limit) was chosen in this simulation.
If $R_0=44 \text{ Gpc}^{-3}\text{yr}^{-1}$ (the upper limit) were used instead, the resulting rates would be increased by a factor of 1.5.

\begin{figure}[h]
	\centering
	\includegraphics[width=0.95\linewidth]{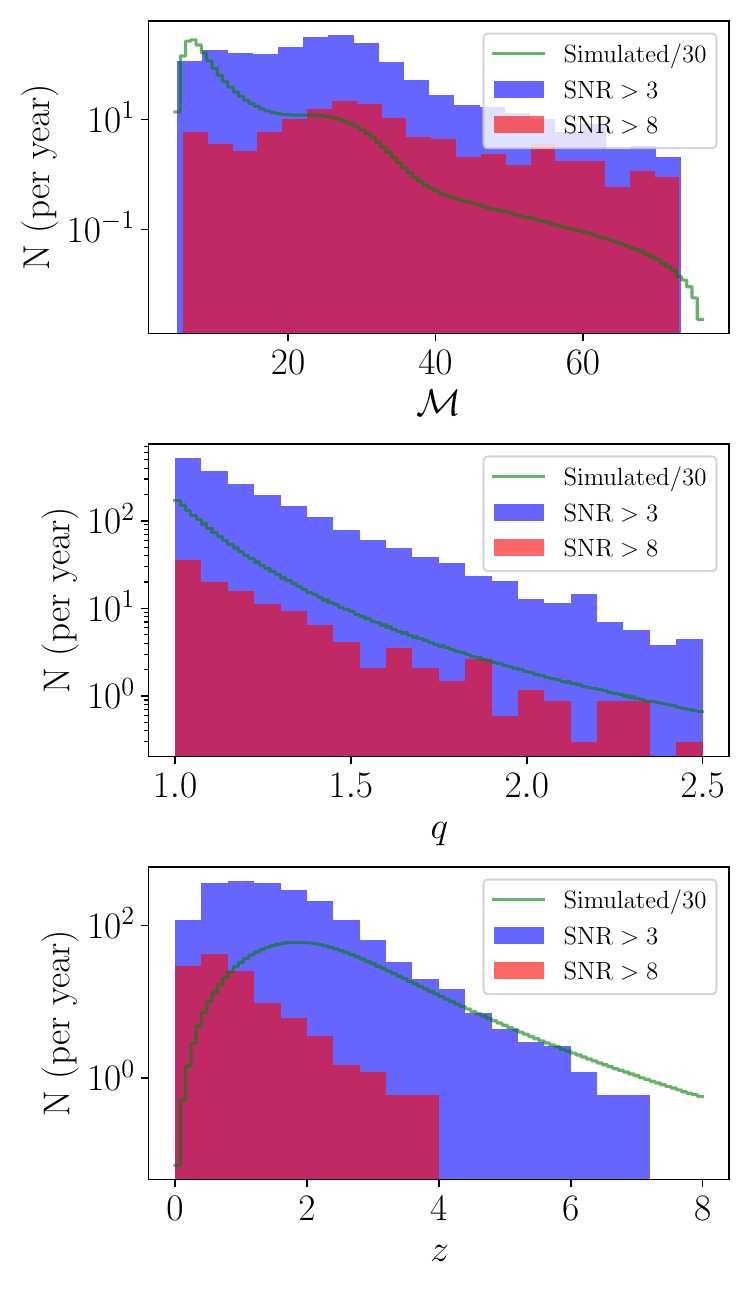}
\caption{\color{black} Distributions of $\mathcal{M}$, $q$, and $z$ for all simulated events (``Simulated/30'', scaled by $1/30$ for readability) and for detectable memory events with $\mathrm{SNR}\geq 3$ and $\mathrm{SNR}\geq 8$.}
	\label{fig-mqz-det}
\end{figure}

{\color{black} Figure~\ref{fig-mqz-det} shows the distributions of $\mathcal{M}$, $q$, and $z$ for the full simulated population (rescaled for readability) together with the subsets detectable with memory SNRs of at least 3 or 8. All panels use identical bin edges (per parameter), and the $y$-axis is in the log scale to enable shape comparisons across thresholds.}

As shown, the observable memory signals for DECIGO would be mainly produced by BBHs with $\mathcal M\sim 30M_\odot$.
A large fraction of these systems have $q \sim 1$.
This agrees with the general feature of the memory signal, i.e., the less the mass ratio is, the larger the memory signal becomes, as mentioned previously.
Although as shown in Fig.~\ref{fig:EventRate}, the chirp mass $\mathcal M$ peaks at $\sim8M_\odot$ for all possible binary star coalescences, here, memory signals produced by binaries with $\mathcal M\sim30M_\odot$ are loudest, because $h_\text{D}$ is an increasing function of $\mathcal M$.
If the detection threshold is set to 3, these sources are dominated by BBHs with $0.25 \lesssim z \lesssim 1.5$.
Naturally, if the threshold is 8, only lower redshift ($\lesssim 0.5$) sources produce sufficiently strong memory signals.
The redshift distributions suggest the robustness of our predictions, as the merger rate provided by the LVK is reliable for $z\lesssim1.5$.

\subsection{The impacts of eccentricities}

Since the LVK collaboration constrained the population model of BBHs assuming the quasi-circular orbits \cite{KAGRA:2021duu},  the simulated merger events in this work have the vanishing eccentricity.
However, the orbit is generally elliptical owing to various astrophysical mechanisms \cite{Rodriguez:2018pss}, and the eccentricity might be large in the bandwidth of DECIGO.
\textcolor{black}{The eccentric orbit influences the evolution of the BBH, and the waveform is modified accordingly.
	Nevertheless, the memory SNR is insensitive to the eccentricity.}

To investigate the influence of the eccentricity on the SNR, we used SEOBNRE \cite{Cao:2017ndf,Liu:2021pkr} to simulate $h_{2,\pm 2}$, assuming $m_1 = m_2 =30 M_\odot$, $D_L=400$ Mpc, $\iota=\pi/2$ and vanishing spins.
One has to set the eccentricity at 10 Hz following the usage of SEOBNRE, although it is capable of calculating the waveform starting at decihertz.
The eccentricities were chosen to be 0.3, 0.2, 0.1, 0.05, 0.01, and 0 at 10 Hz, which correspond to 0.99, 0.98, 0.92, 0.72, 0.15, and 0, respectively, about $5\times10^3$ seconds before the merger.
Then with Eq.~\eqref{eq-hd} to get $h_\text{D}$ and further performing the Fourier transformation, one found out that the SNRs are about 46 to 47 and vary by only about 2\%, showing insignificant impacts of eccentricities on SNRs.
Since we estimated the memory rates based on the SNR, our results are still robust.

\section{Discussions and conclusion}

According to Table~\ref{tab:Results}, DECIGO will detect over $1.5\times10^4$ BBH merger events per year with memory SNRs $>1$.
Such a substantial number surpasses all previous predictions for any detector \cite{Hubner:2019sly,Grant:2022bla,Sun:2022pvh,Gasparotto:2023fcg,Inchauspe:2024ibs}, highlighting the exceptional capability of using the space-borne interferometer to detect an extensive population of stellar-mass BBHs, whose existence has been confirmed by the LVK collaboration.
By concentrating on these sources, our work significantly distinguishes itself from studies involving (super)massive BBHs for LISA and TianQin \cite{Sun:2022pvh,Gasparotto:2023fcg,Inchauspe:2024ibs}  — especially considering that the final parsec problem remains unresolved \cite{Milosavljevic:2002ht}.
Additionally, due to the limited sensitivities at $O(10)$ Hz and below, the LVK detectors might miss some BBH merger events, which could be captured by DECIGO.
Therefore, we might provide the lower bounds.
The idea of detecting memory signals generated by relatively smaller binary stars shall be applied to other space-borne interferometers.
One expects to find some loud memory signals from binaries of  $\sim 10^4 M_\odot$ with LISA, Taiji and TianQin.

The detection of numerous memory signals offers new opportunities for fundamental physics and astrophysics through statistical methods.
For example, global waveform fitting can help examine the sky distribution of memory events, useful for identifying anisotropies in memory SNR.
Also, exploring correlations between memory strength and other parameters, such as mass and spin, can provide insights into nonlinear gravitational dynamics.
The asymptotic symmetries might be better pinned down by analyzing these memory signals statistically \cite{Goncharov:2023woe}.
Moreover, different waveform models can be crosschecked based on the flux-balance laws \cite{Khera:2020mcz}, also using certain statistical methods.

Since a large \textcolor{black}{number} of memory signals come from BBHs with $q \sim 1$ and no precession, the degeneracy between $\iota$ and $D_L$ can thus be broken, which cannot be achieved with the higher multipole modes for such systems \cite{Xu:2024ybt}.
This allows for a more accurate determination of $D_L$, and if $z$ can be obtained, such BBHs would make excellent dark sirens \cite{1987Natur.327..123B,Yang:2022tig}, which are essential for cosmological studies, including the measurement of the Hubble constant.
Although the detection rates for BNS and NS-BH events were not calculated, they are also expected to be quite high, compared to other interferometers.
Memory signals could help distinguish these systems, as suggested in Refs.~\cite{Tiwari:2021gfl,Lopez:2023aja}, improving the estimation of merger rates for different binary types and enhancing our understanding of their formation.

As mentioned, 5 events' memory SNRs, accumulated \textit{solely} in the inspiral phase, are $>3$ .
These events are crucial for testing the memory effect predicted in modified theories of gravity, as nearly all waveforms in these theories have been, or could be, calculated only for the inspiral stage  \cite{Lang:2013fna,Lang:2014osa,Hou:2020tnd,Tahura:2021hbk,Hou:2021oxe,Heisenberg:2023prj,Hou:2024exz}.
Since dipole radiation generally exists in these theories and increases more rapidly at lower frequencies \cite{Chamberlain:2017fjl,Tahura:2021hbk,Heisenberg:2024cjk}, its contribution to the memory waveform may be more easily detected by DECIGO.
Similarly, certain higher-dimensional models could be better constrained or even ruled out.

While next-generation ground-based interferometers like Einstein Telescope (ET) and CE also focus on stellar-mass BBHs, their performance at low frequencies is not as advantageous as that of DECIGO.
However, ET and CE could join with DECIGO if they operate concurrently.
It would enhance the scientific returns of GW observations, allowing for cross-validation of detections and improved parameter estimation.
We leave the detailed exploration of such synergistic observations for future work.

One thus expects that the high detection rates provided by DECIGO and other interferometers would eventually make the study of the memory effect to be more practical, ceasing to be of merely theoretical interest.

\bigskip

\acknowledgements
Z.C.Z would like to thank Xiaolin Liu for help with SEOBNRE code.
This work was supported by the National Natural Science Foundation of China under grant Nos.~11633001 and 11920101003, the Key Program of the  National Natural Science Foundation of China under grant No.~12433001, and the Strategic Priority Research Program of the Chinese Academy of Sciences, grant No.~XDB23000000.
S. H. was supported by the National Natural Science Foundation of China under Grant No.~12205222.
Z.C.Z. was supported by the National Key Research and Development Program of China Grant No. 2021YFC2203001.
{\color{black}
\section{Appendix A:}

Figure~\ref{fig:placeholder} tests the equivalence between constructing the memory directly in the frequency domain and taking a discrete Fourier transform (FFT) of the corresponding time-domain memory built from the same physical inputs. The two spectra are indistinguishable within plotting accuracy over the astrophysically relevant band. Both display the expected low-frequency behavior $|h_{20}(f)| \propto 1/f$, with the usual zero-frequency offset understood to be instrumentally filtered, and any visible departures appear only at the top end of the plot where finite-duration/windowing of the time series affects the FFT. This supports the statement that, at the level of the continuous transform, the two approaches are mathematically equivalent, while direct frequency-domain modeling avoids window-induced ringing.

For contrast, Ref.~\cite{Elhashash:2025hqi} developed time- and frequency-domain models for the dominant $(\ell=2, m=0)$ memory of nonspinning, quasicircular BBHs, calibrated for mass ratios $q\in[1,8]$, and reported mismatches of $O(10^{-2}-10^{-4})$ against NR-surrogate-derived memory with the aLIGO O4 curve. Our approach is complementary: it works directly in the frequency domain and is not restricted to nonspinning systems, allowing to generate waveforms for generic spins and additional modes.

\begin{figure}[h]
    \centering
    \includegraphics[width=1\linewidth]{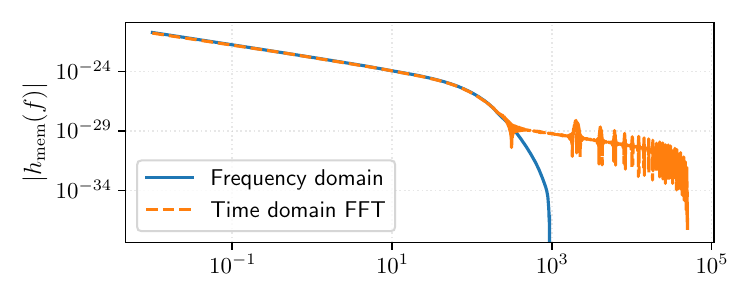}
    \caption{\color{black}Equivalence between our frequency-domain memory model and the FFT of the time-domain memory.}
    \label{fig:placeholder}
\end{figure}
}

\providecommand{\newblock}{}

\end{document}